\documentclass[preprint,showpacs,preprintnumbers,amsmath,amssymb]{revtex4}
\usepackage{graphicx}% Include figure files
\usepackage{dcolumn}% Align table columns on decimal point
\usepackage{bm}% bold math

\begin{document}

\newcommand{\Vint}{V_{int}}
\newcommand{\Eq}[1]{Eq.~(\ref{#1})}
\newcommand{\cph}[1]{Chem. Phys. {\bf #1}}
\newcommand{\cpl}[1]{Chem. Phys. Lett. {\bf #1}}
\newcommand{\prevb}[1]{Phys. Rev. B{\bf {#1}}}
\newcommand{\prevl}[1]{Phys. Rev. Lett. {\bf {#1}}}
\newcommand{\sm}[1]{Synth. Metals {\bf {#1}}}
\newcommand{\jchemp}[1]{J. Chem. Phys. {\bf {#1}}}
\newcommand{\jphysc}[1]{J. Phys. Chem. {\bf {#1}}}
\newcommand{\jpcb}[1]{{J. Phys. Chem. B {\bf #1}}}
\newcommand{\pcp}{$\pi$-conjugated polymer}
\newcommand{\pcs}{$\pi$-conjugated system}
\newcommand{\w}{\mbox{$\omega$}}
\newcommand{\D}{\mbox{$\Delta$}}
\newcommand{\smq}{\mbox{$\simeq$}}
\newcommand{\ee}{E. Ehrenfreund}
\newcommand{\zvv}{Z.V. Vardeny}
\newcommand{\bi}{\bibitem}
\newcommand{\beq}{\begin{equation}}
\newcommand{\eeq}{\end{equation}}
\newcommand{\beqar}{\begin{eqnarray}}
\newcommand{\eeqar}{\end{eqnarray}}
\newcommand{\technion}{Technion--Israel Institute of Technology,
Haifa 32000, Israel}
\newcommand{\phy}{Department of Physics}
\newcommand{\hs}{\hspace{0.4cm}} %standard horizontal space
\newcommand{\npi}{\hspace{-0.4cm}} %no paragraph indent
\newcommand{\vs}{\vspace{0.4cm}} %standard vertical space

\title{
 Negative capacitance in organic semiconductor devices: bipolar
injection and charge recombination mechanism}

\author{
 \ee}
 \affiliation{ Linz Institute for Organic Solar
Cells (LIOS) Johannes Kepler University, Altenbergerstrasse 69, 4040
Linz, Austria, and \phy\ \technion}
 \author {C. Lungenschmied, G. Dennler\footnote{Currently at: Konarka Austria GmbH,
Altenbergerstrasse 69, 4040 Linz, Austria},   H. Neugebauer, N. S.
Sariciftci}
 \affiliation{Linz Institute for Organic Solar
Cells (LIOS) Johannes Kepler University, Altenbergerstrasse 69,
4040 Linz, Austria}
%\author{G. Dennler}
%\footnotetext[]{Currently at: Konarka Austria GmbH,
%Altenbergerstrasse 69, 4040 Linz, Austria }
\begin{abstract}
%\textbf{abstract}

 We report negative capacitance at low frequencies in organic semiconductor based diodes and
 show that it appears only under bipolar injection conditions. We
 account quantitatively for this phenomenon by the recombination current
 due to electron-hole annihilation.
  Simple addition of the recombination current to the well
 established model of space charge limited current in the presence of traps,
 yields excellent fits to the experimentally measured admittance data. The dependence of the
  extracted characteristic recombination time on the bias voltage
 is indicative of a recombination process which is
  mediated by localized traps.

\end{abstract}
 {}

 \maketitle
 \setlength{\columnsep}{6mm}

%\section{\small INTRODUCTION}
 The study and understanding of the optical and electronic properties
of organic light-emitting diodes (OLEDs) is of general importance
and interest, due to their potential as alternatives for classic
inorganic LEDs. The light emitted by LEDs is generated by
electroluminescence (EL), thus detailed knowledge of bipolar
injection and electron hole (e-h) recombination is needed in order
to optimize their device performance. Admittance spectroscopy allows
to differentiate between transport and relaxation processes that
manifest themselves on various time scales. The emphasis in this
work is on the low frequency ($\w$) regime, where a negative
contribution to the capacitance, $C$, has been observed in some
devices.
  Under strong bias conditions, the negative
  contribution dominates at low frequencies and  $C(\w)$ becomes even negative
  \cite{mhb00,hsb03,gkap04,gkj05,psrh05,cbgn05,bgpb06,jasp05}.
 In our devices based on an active layer of a poly(phenylene vinylene) (PPV) derivative, this
negative capacitance (NC) phenomenon occurs only when charges of
both polarity are injected.
 We show that the recombination current leads to NC and derive a
simple expression for its frequency dependence. The NC dependence on
the bias voltage, $V_b$, is inconsistent with Langevin-type
bimolecular recombination, indicating that the e-h recombination is
mediated by localized traps.

The diode devices used in this study were prepared in sandwich
geometry based on structured and cleaned indium tin oxide (ITO)
covered glass substrates (Merck, Inc.).  A 70 nm thick layer of
poly(3,4-ethylenedioxythiophene) doped with poly(styrenesulfonate)
(PEDOT:PSS) (BAYTRON-PH, used as purchased from H. C. Starck) was
applied on the ITO by spin coating. After drying a
poly[2-methoxy-5-(3',7'-dimethyloctyloxy)-1-4-phenylene vinylene]
(MDMO-PPV, from COVION) film was cast from chlorobenzene yielding
a 130 nm thick active layer. Al ("device A") or LiF/Al ("device
B") top electrodes were deposited via thermal evaporation. The LiF
layer has a nominal thickness of 0.7 nm. While a considerable
barrier for electron injection exists between Al and MDMO-PPV, an
almost ohmic contact is expected for LiF/Al \cite{ldens06}. The
complex admittance, $Y(\w)$, is measured using the HP 4192A
impedance analyzer operated in the autobalance mode,
with fixed ac amplitude $v_{ac}(\w)=0.2~V$.\\

  Figure 1  shows
a series of measurements of the capacitance $C\equiv Im(Y/\w)$ vs.
$f (=2\pi/\w)$ from $f$=1 kHz to 1 MHz at various bias voltages
($V_{b}=$ +3 V to +7 V) for device A (top electrode: Al).
 For $V_{b}<$7 V, the main characteristics
are: (a) A knee like feature due to the effect of transit time,
$\tau_t$, is observed at $f\smq 1/\tau_t$ \cite{rk75,mbb99}; (b) At
lower frequencies, $C$ is strongly increasing with decreasing $f$
due to trapping \cite{rk75,mbb99}.
 The average charge carrier mobility is derived from the extracted transit time
 using \cite{lm70}
$\mu_{dc}$=${4L^3}/{3\tau_t E}$, where $E$ is the electric field.
The linear dependence of $log(\mu_{dc})$ on $\sqrt{E}$  (Fig. 1,
inset) is characteristic to polymer devices \cite{mbb99,bb02}.
 Also seen in Fig. 1 is a negative contribution to $C(f)$ at  $f$$\leq$5
kHz for $V_{b}=$7 V (and higher bias).

In order to study the negative contribution to $C(f)$ on a broader
frequency range we utilized device B in which the LiF/Al electrode
allows easier injection of electrons
 into the active layer. In Fig.~2b we show the
electroabsorption (EA) magnitude ($|\D T/T|$) vs. $V_{b}$. The zero
crossing at $V_{b}$$\simeq$+1.8 V yields the internal field
$V_{int}/L$. The actual field across the active layer is therefore
$E=(V_{b}-V_{int})/L$. Fig.~2a shows that the EL due to e-h
radiative recombination sets on just at $V_{b}\simeq V_{int}$,
proving the occurrence of bipolar injection already at this bias.
The measurements of C(f=40 Hz) in the same voltage range (Fig.~2c)
 show that the negative contribution to $C$
starts concurrently with the bipolar injection. At higher $V_{b}$
the negative contribution overwhelms and $C$ becomes negative. The
results for C(f) vs. f in device B (LiF/Al top electrode)  for
various applied bias voltages are displayed in Fig.~3. In the
bipolar injection regime ($V_{b}$$>$+1.8 V) the negative
contribution to $C(f)$ becomes increasingly important as the
frequency decreases. At $V_{b}$$\simeq$+10 V the negative
contribution dominates and $C(f)$ becomes negative below $f\smq4$
kHz. Based on Figs.~2 and 3 we conclude that NC is due to the
presence of charges of both polarities within the active layer of
the device.

%\section{Discussion}
We first summarize previous models that describe the frequency
dependent single carrier transport in low mobility organic diodes.
Assuming a SCLC device with traps, the admittance is written as
\cite{mbb99},
  \beqar
  Y_{SC}(\w)
  =\frac{C_g}{\tau_t}\{\frac{(\w \tau_t)^3}{2i\tilde{\mu}^2[1-\exp{(-i\w \tau_t/\tilde{\mu})}]
  +2\tilde{\mu}\w \tau_t-i(\w \tau_t)^2}\}~, \label{eq1}
   \eeqar
  where $C_g$ is the geometrical capacitance and
    $\tilde{\mu}$ is a normalized dimensionless mobility defined
   as:   $\tilde{\mu}=\mu(\w)/\mu_{dc}$, where $\mu(\w)$ is
   the frequency dependent mobility. In media governed by
   dispersive transport, we use the expression \cite{mbb99}: $\tilde{\mu}(\w)=1+M(i\w \tau_t)^{1-\alpha}$,
   where $\alpha$ is the dispersive exponent and $M$ is a proportionality
   constant.
  The capacitance, $C_{SC}(\w)=Im(Y_{SC}(\w)/\w)$, calculated using \Eq{eq1},
    is then used to fit the measured dynamic capacitance in Fig. 1
    for $V_{b}$$<$7 V.
   The resulting fits, shown as solid lines through the data
   points, appear to be very good, yielding reasonable
   values for the device parameters: $\mu_{dc}$, $\alpha$ and $M$.
   The parameters values agree very well with previously published
   data \cite{mbb99,bb02}. The observed exponential dependence of the mobility on the
   electric field, $\mu_{dc}=\mu_0{\rm exp}(\beta \sqrt{E})$ (see
   the inset in Fig.~1)
    is also in agreement with previous measurements.
    Thus, the description of our device in terms of SCLC including
   traps is reasonably justified.

   Under the conditions of bipolar injection,  the active
   layer contains both electrons and holes.
   When the e-h recombination rate is finite,
   there exists a finite volume in which electrons and holes overlap,
    resulting in a "recombination current". In response to a small
   voltage step, $\D V$, applied at time t=0 and superimposed on a large dc
  bias, there appears an additional, time dependent, current (denoted hereafter as $j_r(t)$) due to the
  recombination. $j_r(t)$ should be proportional to the
  probability of recombination.
   We expect, then, $j_r(t)$ to
  monotonically
  grow from zero at t=0 and to reach slowly the steady state
  value at $t$=$\infty$.   The related capacitance
   is given by the Fourier decomposition \cite{laux85},
  \beqar
  \D C_r(\w)=\frac{1}{\D V}\int_0^{\infty} j_r(t) cos\w t~dt~.
  \label{eq2}
  \eeqar
   In general, Eq.~\ref{eq2} yields a {\underline {negative}} $\D C_r(\w)$ for $j_r(t)$
  as described above (with $d j_r/dt$$>$$0$ and $d^2 j_r/dt^2$$<$$0$) \cite{ellb98}.
  For example, in a trap mediated monomolecular recombination process, the rate equation for the
  charge density, $n$, is $dn/dt=G-n/\tau_r$, where $G$ is the bias dependent
  generation rate and $\tau_r$ is the recombination time.
  Applying a small step $\D G$ at t=0 superimposed
  on a large fixed $G$, the recombination current is given by $
  j_r(t)=\D n/\tau_r$, where $\D n$ is the additional response due
  to $\D G$. Solving the rate equation we obtain
   $ j_r(t)$$\propto$$[1-exp(-t/\tau_r)]$.
  \Eq{eq2} then yields,
  \beqar
 \D C_r(\w)= -{\chi C_g}/({1+\w^2 \tau_r^2})~. \label{eq3}
 \eeqar
 where $\chi$ is a dimensionless parameter that depends on the volume
 where both electrons and holes overlap and on $\tau_r$.
 We obtained identical frequency dependence for the negative capacitance in
 the case of bimolecular recombination.
 The total capacitance is now given by the
 simple addition: $C(\w)=C_{SC}(\w)+\D C_r(\w)$. In Fig.~3 we show (solid lines) the fits obtained
 using Eqs.~\ref{eq1} and \ref{eq3} for the
  measured capacitance in the bipolar injection regime, for various
  bias voltages. In all cases the fits account very well for the
  frequency dependence of the measured total capacitance, including
  the negative contribution.

 We envision two possible recombination mechanisms: direct
 Langevin type bimolecular recombination and recombination
 mediated by localized traps.
 For the bimolecular case, $\tau_r$ should
 decrease with increasing carrier density and mobility.
 Since the carrier
  density and mobility  %\ref{CfAl})
  increase with bias (Fig.~1) $\tau_r$ should
 decrease with increasing bias voltage. This is not observed:
 the values of $\tau_r$ that we obtained from the fits increase from about 0.5 ms at
  $V_{b}$=+2 V to 1--2 ms at $V_{b}$=+10 V.
   For the case
 of trap mediated recombination, on the other hand, $\tau_r$ should be interpreted as
 the characteristic capture time for positive (negative) charge into
 an already negatively (positively) charged trap. The capture rate
 is not expected to increase with the bias, hence no decrease in
 $\tau_r$ is expected with increasing bias.
  %The values $\tau_r$ that we obtained from the fits increase from about 0.5 ms at
%  $V_{b}$=+2 V to 1--2 ms at $V_{b}$=+10 V.  The longer $\tau_r$ at higher bias is
%  inconsistent with direct bimolecular recombination. On the other
%  hand,  trap mediated recombination is governed by the trap capture rate, which is
%  not expected to increase with the applied bias.
    Furthermore, the dimensionless parameter $\chi$
  increases with the bias voltage. This is expected since at higher
  bias voltage the carrier concentration is higher, thus more traps
  are involved in the recombination process.

 %\section{Summary}
 In summary, we have  shown that NC at low frequencies occurs in organic semiconducting diodes
 under bipolar injection and explained it by
 the e-h recombination. The
 time dependent recombination current
 leads to a negative contribution to the low frequency capacitance.
  By suggesting a
simple and straightforward recombination current model, we have
accounted for $C(f)$ on a wide frequency range,
 including the NC region and deduced the characteristic time constant for the
capture/recombination process. The bias dependence of $\tau_r$ is
inconsistent with a bimolecular recombination process. We thus
 % The
 %increase of the NC weight ($\chi$) and the trap capture time
 %constant ($\tau_r$) with bias supports the
 conjecture that in
 organic bipolar devices the e-h
 recombination is governed by trap-mediated processes.\\

  {\bf Acknowledgments}-- Financial support through the European Commission R \& D Programme
(Euro-PSB, NNE5-2001-00544) as well as the Austrian Foundation for
the Advancement of Science (FWF NANORAC, N00103000) is acknowledged.
EE acknowledges the support of the Israel Science Foundation (ISF
735-04).

\newpage

\bibliographystyle{aip}

%\bibliography{NCAPL}% Produces the bibliography via BibTeX.

 \newpage
 Figure captions\\
 \begin{description}
 \item{\bf Figure 1.}
 C vs. f for device A (Al electron injecting electrode) for
$V_{b}$=+3, +4, +5, +7 V. Empty symbols--measured data; solid
lines--fit to the data using Eq.~\ref{eq1}; for $V_{b}$=+7 V
\Eq{eq3} was used as well. Inset: $\mu_{dc}$ vs. the square root of
the net electric field ($E^{1/2}$), extracted using the transit time
obtained from the fits. The solid line is a linear fit of
$log(\mu_{dc})$ vs. $E^{1/2}$.
 \item{\bf Figure 2.}
 EL (a), EA (b) and C (c) vs. $V_{b}$ for device B (LiF/Al
electron injecting electrode). In (b) the magnitude of EA is
plotted. There is a sign change in EA at \smq1.8 V, above which
bipolar injection sets on and both EL and negative contribution to C
are apparent.
 \item{\bf Figure 3.}
  Frequency dependence of the capacitance for device B
(LiF/Al electron injecting electrode) at $V_{b}$=+10 V (2-d plot)
and various other bias voltages (3-d inset). Empty symbols--measured
data; solid lines--fit to the data using Eqs.~\ref{eq1}, \ref{eq3}.
 \end{description}

%\newpage
%  \begin{figure}[h]
%\begin{center}
%\includegraphics[width=12cm]{CvsfAL.eps}
%\caption{C vs. f for device A (Al electron injecting electrode) for
%$V_{b}$=+3, +4, +5, +7 V. Empty symbols--measured data; solid
%lines--fit to the data using Eq.~\ref{eq1}; for $V_{b}$=+7 V
%\Eq{eq3} was used as well. Inset: $\mu_{dc}$ vs. the square root of
%the net electric field ($E^{1/2}$), extracted using the transit time
%obtained from the fits. The solid line is a linear fit of
%$log(\mu_{dc})$ vs. $E^{1/2}$.} \label{CfAl}
%\end{center}
%\end{figure}
%
%\newpage
%  \begin{figure}[h]
%\begin{center}
%\includegraphics[width=12cm]{ELEAC2023.eps}
%\caption{EL (a), EA (b) and C (c) vs. $V_{b}$ for device B (LiF/Al
%electron injecting electrode). Bipolar injection sets on at
%$V_{b}\simeq 1.8 V$, above which both EL and negative contribution
%to C are apparent.}\label{ELEA}
%\end{center}
%\end{figure}
%
%  \newpage
%   \begin{figure}[h]
%\begin{center}
%\includegraphics[width=12cm]{LiFAl2d3d.eps}
%\caption{Frequency dependence of the capacitance for device B
%(LiF/Al electron injecting electrode) at $V_{b}$=+10 V (2-d plot)
%and various other bias voltages (3-d inset). Empty symbols--measured
%data; solid lines--fit to the data using Eqs.~\ref{eq1}, \ref{eq3}.}
%\label{NCLiF}
%\end{center}
%\end{figure}
%

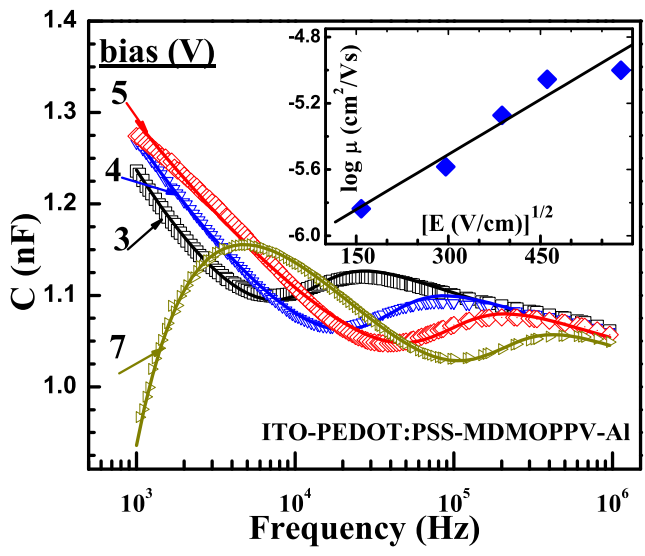
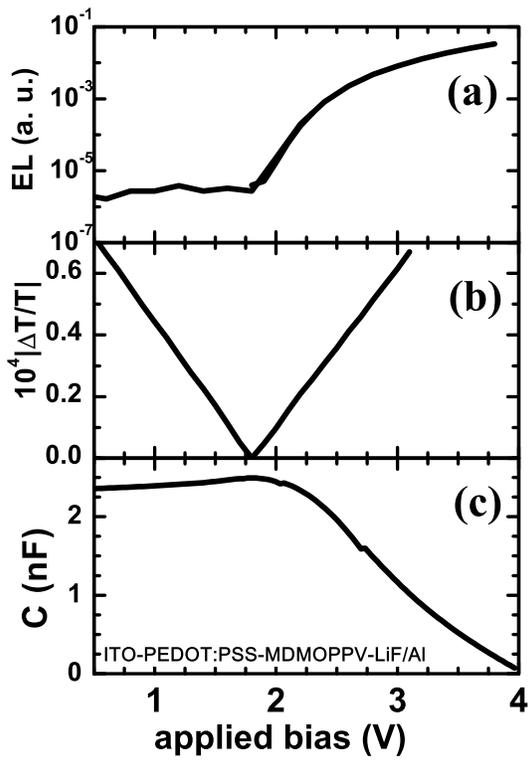
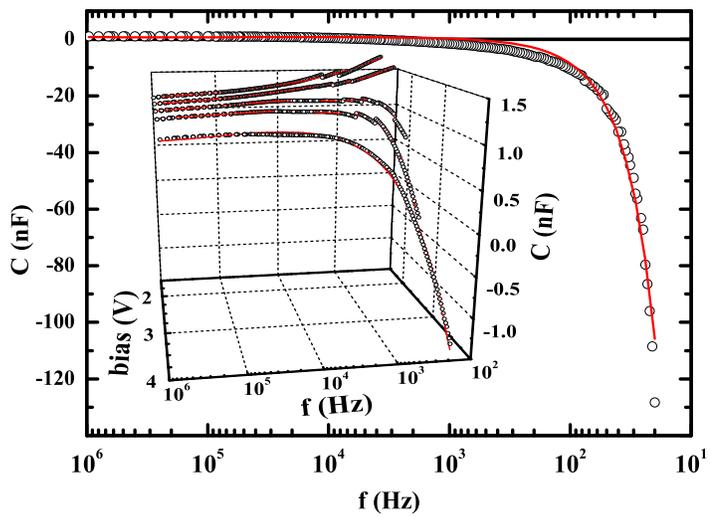
\end{document}